\documentstyle[times]{article}

\begin{document}

\begin{center}
Seeable Matter; Unseeable Antimatter \\
\vspace{.25in}
Geoffrey Dixon \\
gdixon@7stones.com  \\
\vspace{.25in}

\begin{quotation}
The universe we see gives every sign of being composed of matter.  
This is considered a major unsolved problem in theoretical physics.  
Using the mathematical modeling based on the algebra 
${\bf{T}} := {\bf{C}}\otimes{\bf{H}}\otimes{\bf{O}}$, 
an interpretation is developed that suggests that this seeable universe 
is not the whole universe; there is an unseeable part of the universe composed of 
antimatter galaxies and stuff, and an extra 6 dimensions of space  (also 
unseeable) linking the matter side to the antimatter - at the very least.
\end{quotation}
\end{center}
\vspace{.25in}

Bases for the real division algebras, ${\bf{C}}$, ${\bf{H}}$, ${\bf{O}}$ (complex algebra, quaternions, and octonions), are 
\cite{me:bk1}\cite{me:bk2}\cite{7stones}:
$$
\begin{array}{rl}
{\bf{C}} & \{1, \; i\} \\
{\bf{H}} & \{q_{0}=1, \; q_{k}, \; k=1,2,3\} \\
{\bf{O}} & \{e_{0}=1, \; e_{a}, \; a=1,...,7\} \\
\end{array}
$$
The algebra 
$$
{\bf{T}} = {\bf{C}}\otimes{\bf{H}}\otimes{\bf{O}}
$$ 
is $2\times4\times8= 64$-dimensional.  It is noncommutative, 
nonassociative, and nonalternative.

Although I consider it but a restricted model of reality, the basis of what I will do here is the 
10-dimensional space-time model developed in \cite{me:bk1} (chapters 2 to 6), with mathematical expansion 
to be found in \cite{me:bk2} (chapters 2, 3 and 11).  In this model, which accounts for a single family of quarks and 
leptons, and a corresponding antifamily, the foundation is the 128-dimensional hyperspinor space 
$$
{\bf{T}}^{2}
$$
(the doubling of ${\bf{T}}$ in the spinor space is modeled on the notion that a Dirac spinor 
is a double Pauli spinor).

A Dirac spinor is acted upon by the Dirac algebra, 
$$
{\bf{C}}(4) \simeq {\bf{P}}(2),
$$
where the Pauli algebra 
$$
{\bf{P}} \simeq {\bf{C}}(2) \simeq {\bf{C}}\otimes{\bf{H}}.
$$
This is the complexification of the Clifford algebra if 1,3-spacetime.  
Likewise ${\bf{T}}^{2}$ is acted upon by the complexification of the Clifford algebra of 
1,9-spacetime, represented by 
$$
{\bf{T}}_{L}(2),
$$
where ${\bf{T}}_{L}$ is the algebra of left actions of ${\bf{T}}$ on itself, which in the octonion 
case, due to nonassociativity, requires the nesting of actions (see, for example,  \cite{me:bk1}, 
chapter 2, and \cite{me:bk2}, section 2.4; and for more background material, \cite{baez},  
\cite{conway}, and \cite{gursey} (the work of G{\"u}rsey at Yale University during the 1970s 
was the inspiration for all of my work - and that of many others - applying the octonion 
algebra to physics)).  

The work of G{\"u}rsey (and G{\"u}naydin) was inspired by the work of von Neumann, Jordan and 
Wigner \cite{vonneumann},  
who investigated an expansion of quantum theory from a foundation on ${\bf{C}}$ to one on ${\bf{O}}$. 
They linked quantum observability with algebraic associativity, and unobservability with nonassociativity, 
thinking along these lines being forced by the nonassociativity of ${\bf{O}}$.  
(I do not know the details of their work, but the notion that nonassociativity could be associated 
with things unseen, and unseeable, partly motivated this work.)  

The quantum notion of unobservable is more restrictive than the notion of unseeable being used here.  
In particular, quarks are unseeable, but they are detectable, and they supply the paradigm - 
albeit not well defined - of what is meant by unseeable.  But being not well defined is not a problem.  
My working assumption is that this model is but a kernel 
of something much larger, much deeper, and, I hope, ultimately knowable.  It may not be.

The model building in  \cite{me:bk1}\cite{me:bk2} relies heavily on a resolution of the 
identity of 
$$
{\bf{S}} := {\bf{C}}\otimes{\bf{O}}
$$
into a pair of orthogonal idempotents, 
$$
\rho_{\pm} = \frac{1}{2}(1 \pm ie_{7}).
$$
These satisfy 
$$
\rho_{\pm}e_{p}\rho_{\pm} = e_{p}\rho_{\mp}\rho_{\pm} = 0, \; p = 1,2,3,4,5,6,
$$
and
$$
\rho_{\pm}e_{k}\rho_{\pm} = e_{k}\rho_{\pm}\rho_{\pm} = e_{k}\rho_{\pm}, \; k = 0,7
$$
(nonassociativity does not play a role here, so no parentheses are required; also note that 
$e_{7}\rho_{\pm} = \mp i \rho_{\pm}$).
(This is the same resolution exploited by G{\"u}rsey, et al. \cite{gursey}, and numerous other places in the 
years following.  As is done here, it is how they gave rise to the $SU(3)$ color group out of 
the octonion automorphism group, $G_{2}$ (see \cite{me:bk1}, chapter 2).)  
With these projectors ${\bf{S}}$ can be divided into 4 orthogonal subspaces:
$$
\begin{array}{rl}
{\bf{S}}_{++} & = \rho_{+}{\bf{S}}\rho_{+}, \\
{\bf{S}}_{--} & = \rho_{-}{\bf{S}}\rho_{-}, \\
{\bf{S}}_{+-} & = \rho_{+}{\bf{S}}\rho_{-}, \\
{\bf{S}}_{-+} & = \rho_{-}{\bf{S}}\rho_{+}.
\end{array}
$$
Both ${\bf{S}}_{++}$ and ${\bf{S}}_{--}$ are associative subalgebras of ${\bf{S}}$ isomorphic to ${\bf{C}}$.  
${\bf{S}}_{+-}$ and ${\bf{S}}_{-+}$ are not subalgebras, and they are highly nonassociative (this nonassociativity 
implying ${\bf{S}}_{\pm\mp}^{2} = {\bf{S}}_{\mp\pm}$ (you'd better check that - it's not relevant, but I never 
noticed that before - hmm)).  Anyway, elements of the first two sets are linear (over ${\bf{C}}$) in the octonions 
$\{e_{0} = 1, \; e_{7}\}$, and the second two sets linear over $\{e_{p}, \; p = 1,2,3,4,5,6\}$.

With respect to the $SU(3)$ subgroup of the octonion automorphism group $G_{2}$ that leaves 
the unit $e_{7}$ fixed these parts of $\bf{S}$ transform, respectively, as a singlet, anti-singlet, triplet, and anti-triplet.  
That is, 
$$
\begin{array}{l}
\rho_{+}{\bf{S}} \mbox{ is matter;} \\
\rho_{-}{\bf{S}} \mbox{ is anti-matter.}
\end{array}
$$
The same is true if we replace $\bf{S}$ by ${\bf{T}}^{2}$.

An elegant representation of the Clifford algebra ${\cal{CL}}(1,9)$ represented in ${\bf{T}}_{L}(2)$, that is aligned 
with the choice of the octonion unit $e_{7}$ to appear in $\rho_{\pm}$, arises from the following set of ten 
anti-commuting 1-vectors: 
$$
\beta, \,\,\,
\gamma q_{Lk}e_{L7}, \, k = 1,2,3, \,\,\,
\gamma i e_{Lp}, \, p = 1,...,6, 
$$
where
$$
\epsilon = 
\left[\begin{array}{rr}
1 & 0 \\ 0 & 1 \\
\end{array}\right], \,\,
\alpha = 
\left[\begin{array}{rr}
1 & 0 \\ 0 & -1 \\
\end{array}\right], \,\,
\beta = 
\left[\begin{array}{rr}
0 & 1 \\ 1 & 0 \\
\end{array}\right], \,\,
\gamma = 
\left[\begin{array}{rr}
0 & 1 \\ -1 & 0 \\
\end{array}\right],
$$
and as usual the subscripts "L" and "R" signify an action from the left or the right on ${\bf{T}}$.  (So, for example, 
$$
{\bf{S}}_{+-} = \rho_{+}{\bf{S}}\rho_{-} = \rho_{L+}\rho_{R-}[{\bf{S}}].)
$$
(Note: ${\cal{CL}}(1,9)$ can be represented in other ways, and certainly using only complex matrices.  
A principal underlying this work is that the division algebras should be as generative in physics as they 
are in mathematics, and that, with the Dirac algebra and its spinors as a guide, this model of 
1,9-spacetime, with its spinor space consisting of a family and antifamily of quark and lepton 
Dirac spinors, falls out relatively naturally, if one pays attention to the structure of the underlying 
mathematics.  The alignment of this representation of ${\cal{CL}}(1,9)$ with the $\rho$ projectors 
is not necessary, but then neither is a combination necessary to open a safe.  Dynamite will do.  
By ordering things as I have, I am attempting to demonstrate the elegant way the mathematics 
elucidates the physics - to provide a combination with which the goodies in this safe can be more easily 
grasped, and the essential nature of the mathematics made more clear.)

Here is the working assumption upon which this work is based: if we project out from this model 
those bits we know are unseeable (anything carrying a color charge), what is left will be seeable, 
and everything that is gone will be unseeable (even if it does not carry a color charge).  
As it stands, this is the model of a universe with 10 dimensions, containing both 
matter and anti-matter in the form of leptons and quarks, and their anti-particles.  The quarks and 
the extra 6 space dimensions are unseeable.   There exist models that attempt to explain quark 
confinement, but so long as they remain confined we are safe in labeling them unseeable.  

Quarks carry $SU(3)$ color charges, as do the extra 6 spaces dimensions in this model.  Both are 
unseeable (this is both an assumption, and an observation), and both can be projected out of the 
model in the same way.  Since the color charges 
reside in the octonion units $e_{p}$, p = 1,...,6, we need merely use the $\rho_{\pm}$ to get rid of 
them.

Start with the 6 extra space dimensions.  There are two (what I would call) canonical 
ways of reducing the 1-vectors of ${\cal{CL}}(1,9)$, a mix of seeable and unseeable dimensions, to the 
1-vectors of seeable ${\cal{CL}}(1,3)$ (that is, we are using the $\rho$ projectors to eliminate bits 
that carry the unseeable color charge, here the extra 6 space dimensions):
$$
\begin{array}{c}
\rho_{L\pm}\; \{\beta, \,\,\, \gamma q_{Lk}e_{L7}, \, k = 1,2,3, \,\,\, \gamma i e_{Lp}, \, p = 1,...,6 \}\rho_{L\pm} \\ \\
= \;  \{\beta, \,\,\, \gamma iq_{Lk}, \, k = 1,2,3 \}\rho_{L\pm}.
\end{array}
$$
These two collections of ${\cal{CL}}(1,3)$ 1-vectors act on half of the full spinor space ${\bf{T}}^{2}$.  In particular, 
they act respectively on 
$$
\rho_{L\pm}[{\bf{T}}^{2}] = \rho_{\pm}{\bf{T}}^{2},
$$
where the underlying mathematics implies that these are, respectively, the matter and anti-matter halves of ${\bf{T}}^{2}$ 
($\rho_{+}{\bf{T}}^{2}$ being a full family of lepton and quark Dirac spinors, and $\rho_{-}{\bf{T}}^{2}$ the corresponding 
anti-family: see \cite{me:bk1}, chapters 3 and 4, and \cite{me:bk2}, section 3.2)).

And this is the point: once 1,9-spacetime is reduced to 1,3-spacetime (the unseeable part projected away), 
one discovers that half of the hyper-spinor space is also projected away, and it too - given the interpretation 
of the mathematics we are adopting here - should be unseeable, even though bits of it do not carry the color charge 
(anti-leptons).  That is, from the 1,3-spacetime that is left you can 
see only the matter half of ${\bf{T}}^{2}$, or the antimatter half.  One or the other is projected away, along with 
things carrying the color charge, and so this antimatter universe should too be unseeable.  We think of 
our universe as being composed of matter (stars, planets, and such; the production of individual antimatter 
particles is not considered a problem).  The antimatter half of ${\bf{T}}^{2}$ is not gone, nor are the extra 
6 space dimensions.  We just don't directly see them.

Quarks, like the extra 6 dimensions of space in this model, are also unseeable.  
And like the extra 6 dimensions of space, they owe their existence to the octonion units $e_{p}, \; p=1,2,3,4,5,6$.
To reduce the spinor space ${\bf{T}}^{2}$ all the way to its observable lepton part (the anti-lepton part is similar) we need 
an extra $\rho_{+}$.  Specifically, 
$$
\rho_{L\pm}\rho_{R\pm}[{\bf{T}}^{2}] = \rho_{+}{\bf{T}}^{2}\rho_{+}
$$
is a lepton doublet, consisting of 2 Dirac spinors, one for the electron, one for its neutrino.  (The particle identifications 
are not arbitrary.  See particularly \cite{me:bk2}, section 3.2, for the mathematics behind that statement.)  Interestingly, this further 
reduction does not result in any further reduction of the 1-vector space of our original Clifford algebra, ${\cal{CL}}(1,9)$.  
We're still left with a version of 1-vectors for ${\cal{CL}}(1,3)$.  However, the story is different for the space of 2-vectors.  
Initially they form a representation of the 1,9-Lorentz Lie algebra, $so(1,9)$.  After the initial reduction we get something 
more than $so(1,3)$:
$$
\rho_{L+}so(1,9)\rho_{L+} = (so(1,3) \times so(6)) \rho_{L+},
$$
and after the second spinor reduction, 
$$
\rho_{R+}\rho_{L+}so(1,9)\rho_{L+}\rho_{R+} = (so(1,3) \times u(1) \times su(3)) \rho_{L+}\rho_{R+}.
$$
This is precisely what it seems, and precisely the part of $so(1,9)$ we observe to function in our seeable part of 
the universe.  (Isospin $SU(2)$ arises from ${\bf{H}}_{R}$ (see \cite{me:bk1}, section 3.5; \cite{me:bk2}, chapter 3; and 
\cite{me:jmp} for an extension of these ideas).  In short, ${\bf{H}}_{R}$ is isomorphic to ${\bf{H}}$; the elements of unit norm are 
the 3-sphere, $S^{3} \simeq SU(2)$; and this $SU(2)$  commutes with the Clifford 
algebra for 1,9-spacetime developed above, so it is an internal symmetry with respect to that spacetime.)

The situation is more complicated than this (see \cite{me:bk1}\cite{me:bk2}), but the overriding point being made here is 
that the mathematics of ${\bf{T}}$ can be viewed as implying we exist in an observable universe that must be dominantly 
matter, or anti-matter (if we accept that everything carrying nontrivial $SU(3)$ color charges is not directly observable by us, 
which in this context includes quarks, anti-quarks, and the extra 6 dimensions of spacetime, all of which 
involve the octonion units, $e_{p}, \; p = 1,2,3,4,5,6$, which carry those charges).  Acceptance of this notion has the potential 
to imply far more profound things about physics.  

I consider this an elegant explanation of why we perceive our universe to be composed of matter.  There are 
many (a great many) open questions that will not be resolved here.  Quarks, as mentioned, are unseeable, but detectable.  
This color confinement is thought to be related to energy considerations of the strong force - but confinement it is.  So the 
question arises: are the extra 6 (or more) dimensions of space detectable, and if so, what is the mechanism that hides them 
from us?  Is the antimatter universe detectable, and what mechanism hides it?  
(Note: our observable universe has the occasional antimatter particle whizzing around.  It is not being suggested that 
these should be unseeable, but that there is an antimatter universe out there (whatever "there" means) that we do not see.)  
And beyond this, what is really needed is a (much) deeper theory from which one might glean insights into these unresolved 
problems.

A penultimate note: in \cite{me:bk1} (section 6.3) it was pointed out that the original model allowed algebraically for matter-antimatter mixing 
via the extra 6 dimensions, but that reasonable conditions put on the dependence of the various particle fields on these 
extra dimensions led to these mixing pathways disappearing.  Whatever the case, this idea of mixing is mediated by those 
extra 6 dimensions, which provide channels from the matter part of the overall universe to the antimatter part.  
Were these channels viable they would allow, for example, an electron from our matter part to channel through 
to the antimatter part, appearing on the other side as an antiquark (it necessarily picks up an anti-color charge 
en route).  But this idea just scratches the surface.

And finally, I would like to add that this exploitation of ${\bf{T}}$ as the foundation of a model of reality is not the only one, it is the one I like 
best (well, I've been at it for over 30 years, so changing now is not going to happen).  For an alternate approach, see \cite{cohl}.

\end{document}